\documentclass[preprint, trackchanges]{aastex62}

\usepackage[version=3]{mhchem}
\usepackage{chemfig}
\tabletypesize{\footnotesize}

\turnoffeditone

\received{}
\revised{}
\accepted{}

\shorttitle{}
\shortauthors{Iino et al. 2021a}

\begin{document}

\title{$^{13}$C isotopic ratios of \ce{HC3N} on Titan measured with ALMA}

\author{Takahiro Iino}
\affil{Information Technology Center, The University of Tokyo, 
2-11-16 Yayoi, Bunkyo, Tokyo 113-8658, Japan}
\email{iino@nagoya-u.jp}

\author{Kotomi Taniguchi}
\affil{National Astronomical Observatory of Japan, 2-21-1 Osawa, Mitaka, Tokyo 181-8588, Japan}

\author{Hideo Sagawa}
\affil{Faculty of Science, Kyoto Sangyo University, Motoyama, Kamigamo, Kita-ku, Kyoto 603-8555, Japan}

\author{Takashi Tsukagoshi}
\affil{National Astronomical Observatory of Japan, 2-21-1 Osawa, Mitaka, Tokyo 181-8588, Japan}

\begin{abstract}
We present the first determination of the abundance ratios of $^{13}$C substitutions of cyanoacetylene (\ce{HC3N}), [\ce{H^{13}CCCN}]:[\ce{HC^{13}CCN}]:[\ce{HCC^{13}CN}] in Titan's atmosphere measured using millimeter-wave spectra obtained by the Atacama Large Millimeter-submillimeter Array.  
To compare the line intensities precisely, datasets which include multiple molecular lines were extracted to suppress effects of Titan's environmental conditions and observation settings.
The [\ce{HC^{13}CCN}]:[\ce{HCC^{13}CN}] and [\ce{H^{13}CCCN}]:[\ce{HCC^{13}CN}] ratios were obtained from 12 and 1 selected datasets, respectively. 
As a result, nearly the uniform [\ce{H^{13}CCCN}]:[\ce{HC^{13}CCN}]:[\ce{HCC^{13}CN}] abundance ratios as 1.17 ($\pm$0.20) : 1.09 ($\pm$0.25) : 1 (1$\sigma$) were derived, whereas previously reported ratios for interstellar medium (ISM) have shown large anomalies that may be caused by $^{13}$C concentrations in precursors.
The result obtained here suggests that $^{13}$C concentration processes suggested in the ISM studies do not work effectively on precursors of \ce{HC3N} \edit1{and \ce{HC3N} itself } due to Titan's high atmospheric temperature \edit1{ and/or depletion of both \ce{^{13}C} and \ce{^{13}C+}}. 
\end{abstract}


\section{Introduction}
The $^{13}$C substituted species of cyanoacetylene (\ce{HC3N}), namely, \ce{H^{13}CCCN}, \ce{HC^{13}CCN} and \ce{HCC^{13}CN}, have been discovered in various interstellar media (ISM), and are known to exhibit large isotopic anomalies that \ce{HCC^{13}CN} and/or \ce{H^{13}CCCN} show high $^{13}$C concentrations \citep{Takano1998, Taniguchi2016, Araki2016, Taniguchi2017}.
Such anomalies are considered to be due to the $^{13}$C concentrations on the precursors of \ce{HC3N} such as CN and \ce{C2H} \citep{Furuya2011, Taniguchi2019}, and give us important information on the environments and possible chemical reactions in the ISM. 

\ce{HC3N} is also present on Saturn's largest moon, Titan, and was first detected in the atmosphere by the Voyager 1 spacecraft \citep{Kunde1981}.
\edit1{The main production pathway of \ce{HC3N} was expected as follows:
\begin{equation}
    \ce{C2H2 + CN -> HC3N + H} \label{eq:CN}
\end{equation}
}
Since then, a number of in-situ, ground- and space-based observations have been performed to illustrate the spatial and time variation of \ce{HC3N} along with other nitriles and hydrocarbons \citep{Hidayat1997, Coustenis1998, Coustenis2003,Gurwell2004, Coustenis2007, Coustenis2010,Cordiner2014, Coustenis2016, Coustenis2018,  Thelen2019}.

Adding to Reaction \ref{eq:CN}, a following reaction of \ce{C2H} radical with HNC \citep{Loison2015} possibly produces a portion of \ce{HC3N} because HNC is present in the upper stratosphere \citep{Moreno2011, Cordiner2014}.
\begin{equation}
    \ce{C2H + HNC -> HC3N + H} \label{eq:HNC}
\end{equation}
In turn, due to the extremely high reaction barrier, the reaction of \ce{C2H} with HCN, an isomer of HNC, does not work effectively. 
Thus, Reaction \ref{eq:HNC} is the only reaction to produce \ce{HC3N} from \ce{C2H} radical.

\edit1{
Two precursor radicals of \ce{HC3N}, namely CN and \ce{C2H}, are important for the \ce{HC3N} production along with the reaction counterparts, \ce{C2H2} and HNC. 
They are easily produced by the photo-dissociation HCN and \ce{C2H2} molecules, and their expected abundances are $\sim$10 ppb at 1000 km \citep{Lavvas2008}. 
}

As for the ISM, $^{13}$C substitutions of \ce{HC3N} have also been detected on Titan.
The first observational result was reported by \cite{Jennings2008} using infrared spectra obtained with the Composite Infrared Spectrometer on-board Cassini spacecraft. 
Spectral emissions of the three isotopologues were clearly detected, whereas the \ce{HC^{13}CCN} and \ce{HCC^{13}CN} lines were blended. 
Using the \ce{H^{13}CCCN} and \ce{HC3N} lines, the $^{12}$C/$^{13}$C ratio was measured to be 79$\pm$17, which is consistent with that measured in HCN (70--120: \cite{Hidayat1997}, 132$\pm$25 or 108:$\pm$20:\cite{Gurwell2004}, 89.8$\pm$2.8:\cite{Molter2016} ) and \ce{C2H2} (84.8$\pm$3.2:\cite{Nixon2012}). 
A recent submillimeter spectroscopy using ALMA succeeded in the detection of \ce{H^{13}CCCN} with a high S/N ratio along with \ce{HCCC^{15}N} \citep{Cordiner2018}, although the $^{12}$C/$^{13}$C value was not determined for \ce{HC3N}. 

In this study, we report the first observational determination of the relative \ce{^{13}C} carbon isotopic ratios of three isotopologues of \ce{HC3N}, namely  [\ce{H^{13}CCCN}]:[\ce{HC^{13}CCN}]:[\ce{HCC^{13}CN}] on Titan by analyzing a large dataset obtained by ALMA.
The result enables us to compare the chemical environment of Titan with that of the ISM.

\section{Analysis}
\subsection{Data selection}
Since Titan is often used as a calibrator of ALMA, a large amount of observation data of Titan is available in the ALMA archive. 
We calibrated and imaged all of the archived observational data of Titan that were available as of January 2020.
The details of the calibration and imaging procedure were as described in our previous paper that analyzed nitrogen isotopic ratio of \ce{CH3CN} on Titan \citep{Iino2020}.

Spectral lines of the isotopologues of \ce{HC3N} are often observed by ALMA by chance because their pure rotational transitions appear every $\sim$10 GHz.  
To measure the line intensities precisely, we have chosen spectral windows (SPW) that observed multiple isotopologues simultaneously. 
The usage of the data in the same SPW suppresses the systematic uncertainties arising from the differences in the observation configurations such as the synthesized beam size and absolute flux calibration, and Titan's environmental conditions such as the horizontal and vertical distribution of \ce{HC3N} and the atmospheric structure.
To measure the line intensities that have a narrow ($\sim$1.5 MHz) line-width, SPW that have a high frequency resolution of $<$2 MHz were chosen.
The frequency difference between \ce{HC^{13}CCN} and \ce{HCC^{13}CN} that share the same rotational state $J$ is no more than 20 MHz because they have a very similar rotational constant $B$.
Thus, in most cases, they are observed in the same SPW.
In turn, since the rotational constant of \ce{H^{13}CCCN} is $\sim$2$\%$ smaller than those of \ce{HC^{13}CCN} and \ce{HCC^{13}CN}, the number of SPW including three lines was smaller than that includes \ce{HC^{13}CCN} - \ce{HCC^{13}CN} pair.  

\edit1{
As an important phenomenon, Titan’s trace gases including \ce{HC3N} and its precursors, \ce{C2H2} and HCN, are known to exhibit large spatial and time variations. 
The analyzed period, from 2012 to 2015, is a season of northern summer, when increase and decrease of trace species have been observed by Cassini and ALMA for southern and northern hemispheres, respectively  \citep{Cordiner2015, Thelen2019, Cordiner2019, Coustenis2018}. 
}

\edit1{
To decrease the effect of such data-to-data variability of the \ce{HC3N} spatial distribution, disk-averaged spectra were extracted from the imaged cube fits with a 0.$\arcsec$4 radius circle which is large enough to cover the entire disk of Titan for all of the analyzed datasets.
}

The baseline structure \edit1{of the spectra} was attempted to be removed using polynomial fitting method, while the effect was very limited. 
For the line intensity measurement, spectral intensities within a range of $\pm$1 MHz from the line center were integrated. 
\edit1{Noise level was measured in the line-free region and multiplied by $\sqrt{n}$ where n is the number of averaged channel.}
After extracting 29 SPW which include multiple emission lines, 
SPW that exhibit high S/N ratio as $>$4 were chosen for the intensity ratio measurement analysis.

The number of selected SPW including \ce{HC^{13}CCN} - \ce{HCC^{13}CN} and \ce{H^{13}CCCN} - \ce{HCC^{13}CN} pairs obtained with high S/N ratio were 12 and 1, respectively. 
Observation parameters of the selected SPW are summarized in Table \ref{table:observation}.
The rotational state transitions corresponding to 217, 226, 235, 244 and 271 GHz bands were $J$ = 24--23, 25--24, 26--25, 27--26 and 30--29, respectively. 
For an selected \ce{H^{13}CCCN} - \ce{HCC^{13}CN} pair, they have different transition of $J$=35--34 and 34--33 for \ce{H^{13}CCCN} and \ce{HCC^{13}CN}, respectively. 
Most of the bands are of the ALMA Band 6, except for 308 GHz of the Band 7.
The project code 2015.1.00512.S data has a long observation time by concatenating short observation time data to improve the S/N ratio. 

\begin{figure*}
\begin{center}
\includegraphics[scale=0.3]{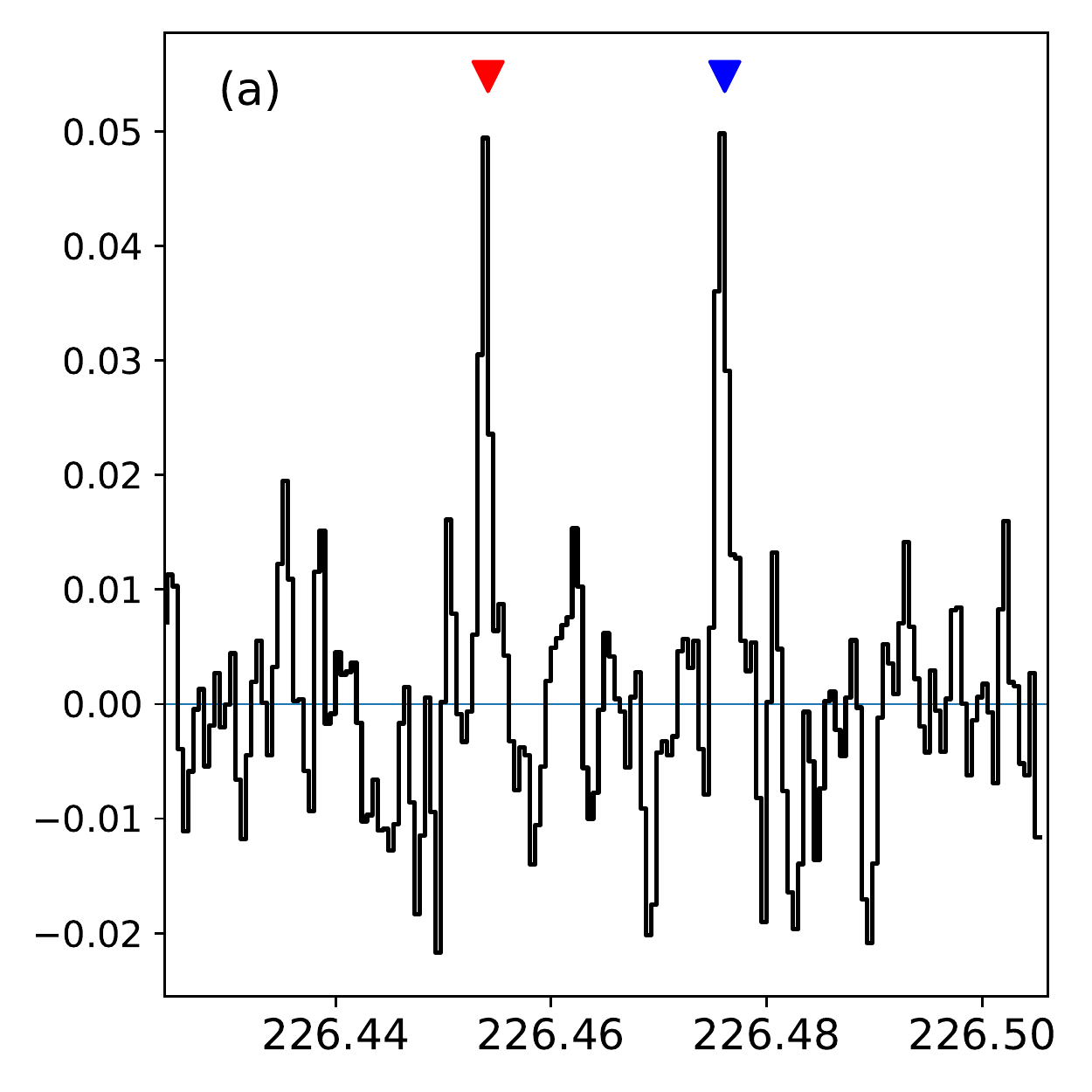}
\includegraphics[scale=0.3]{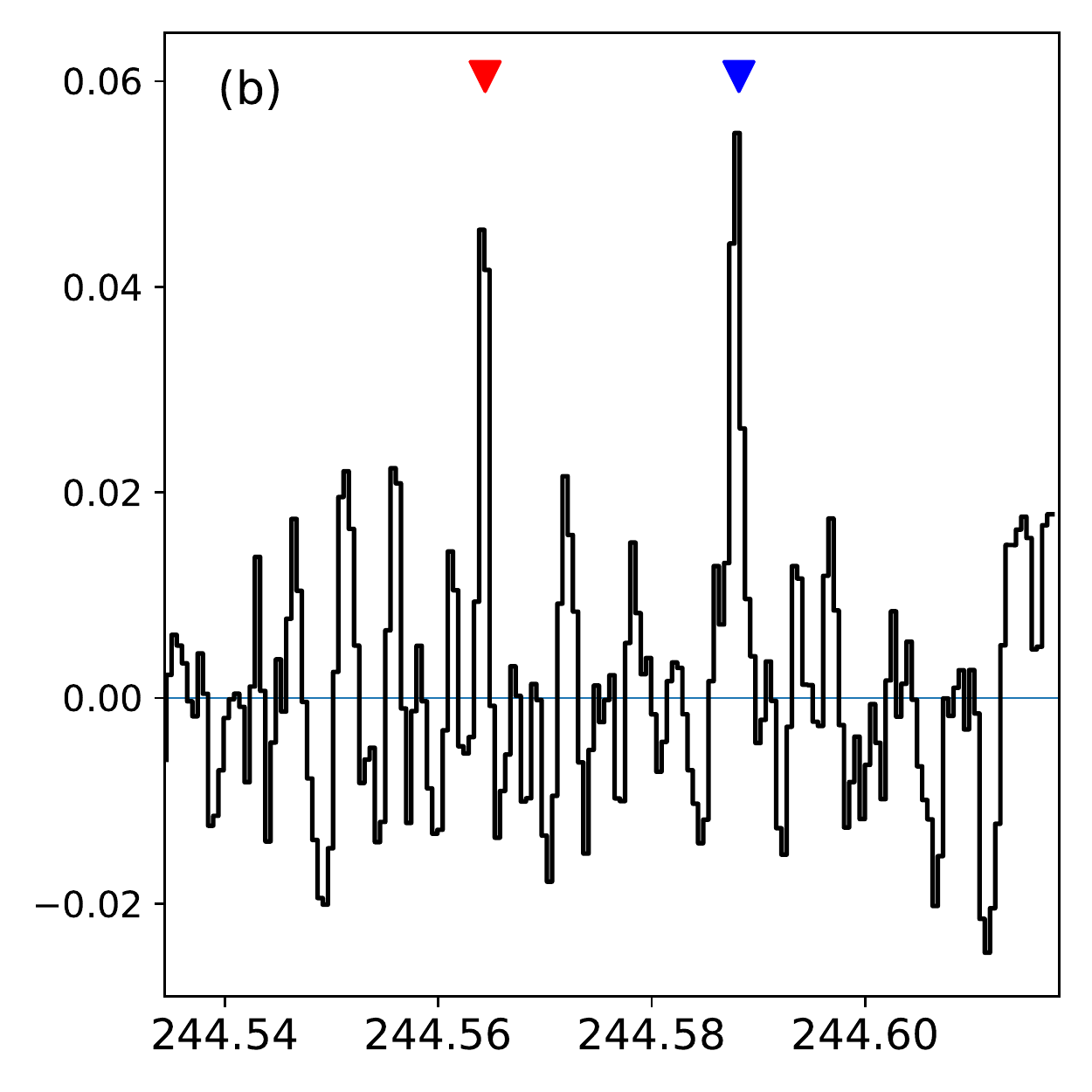}
\includegraphics[scale=0.3]{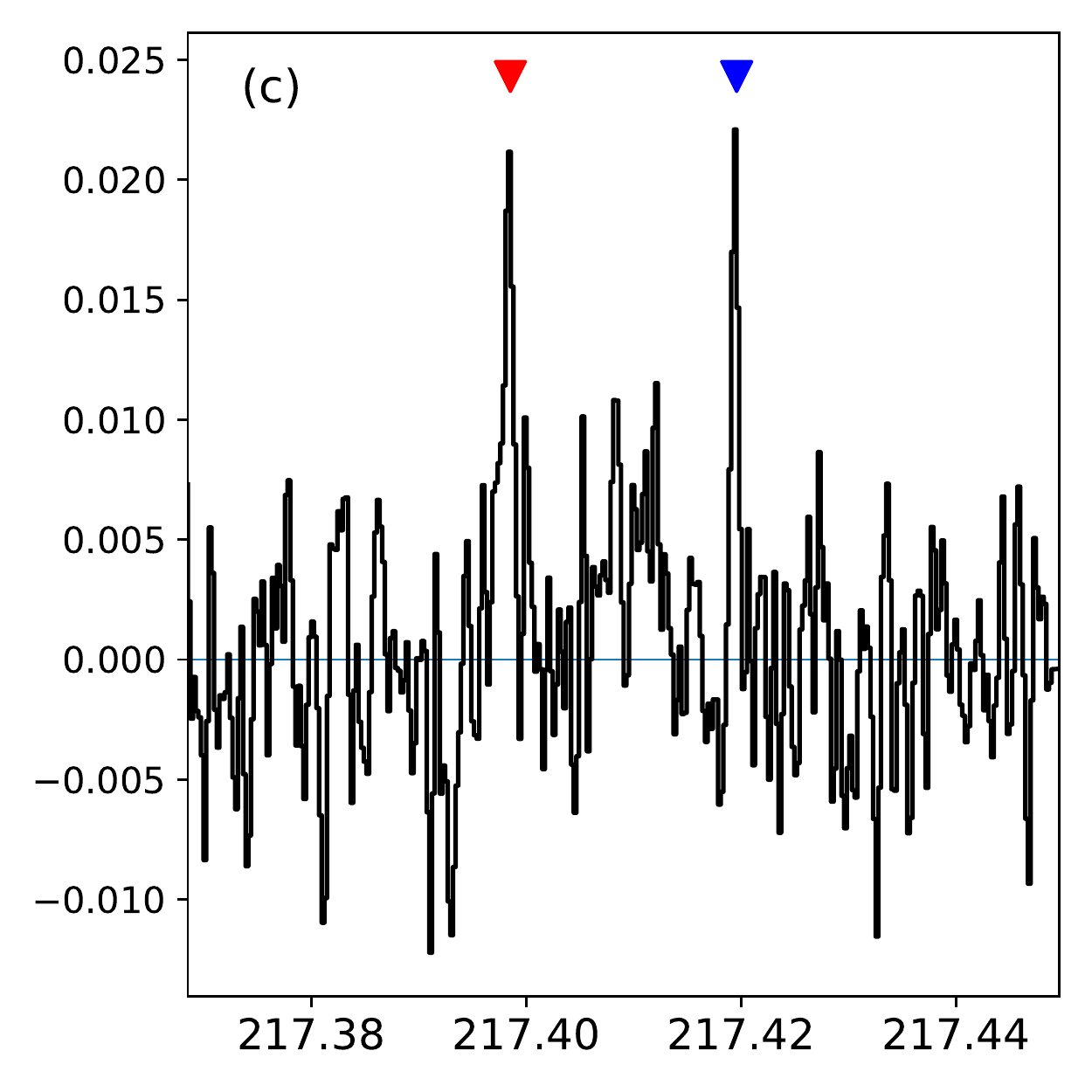}
\includegraphics[scale=0.3]{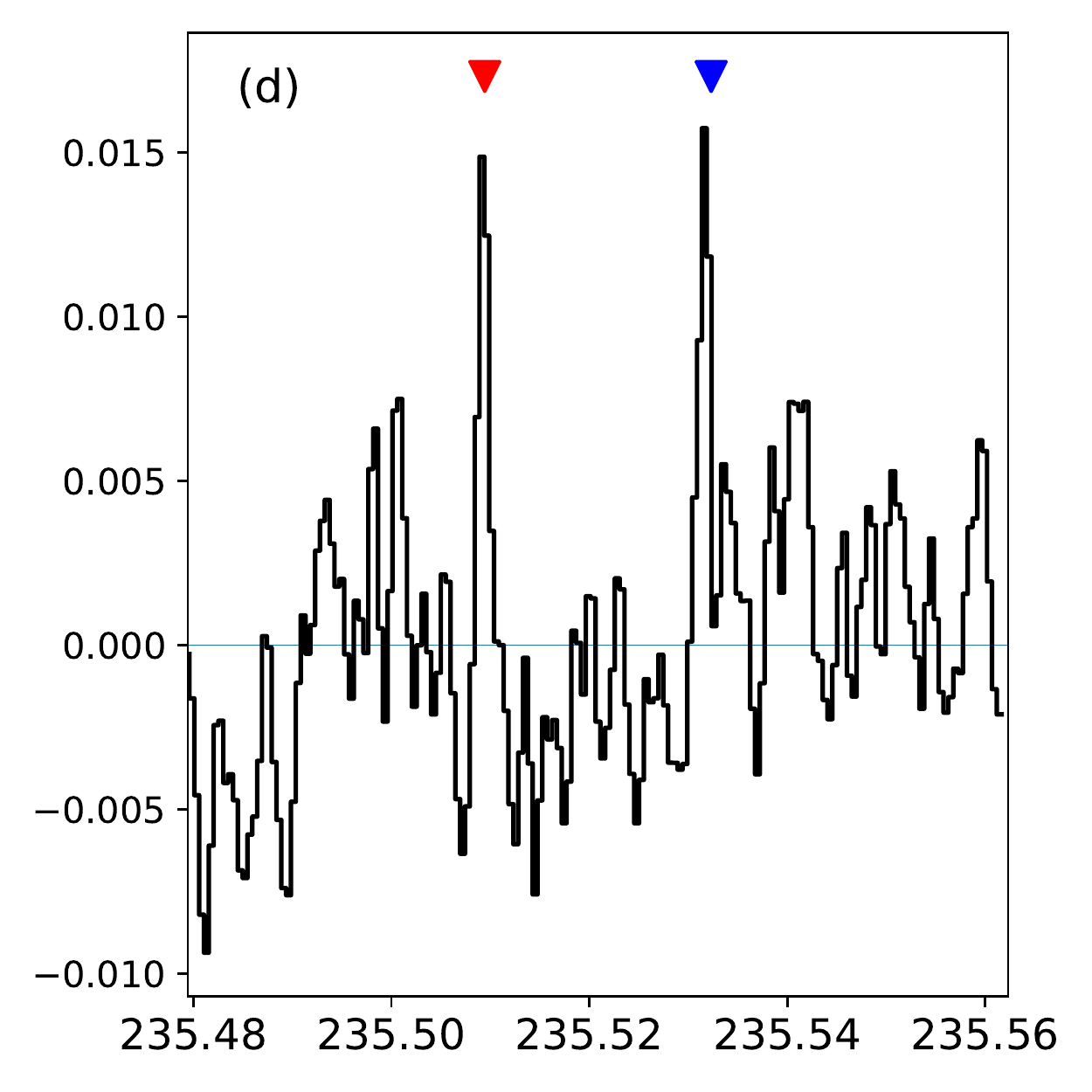}
\includegraphics[scale=0.3]{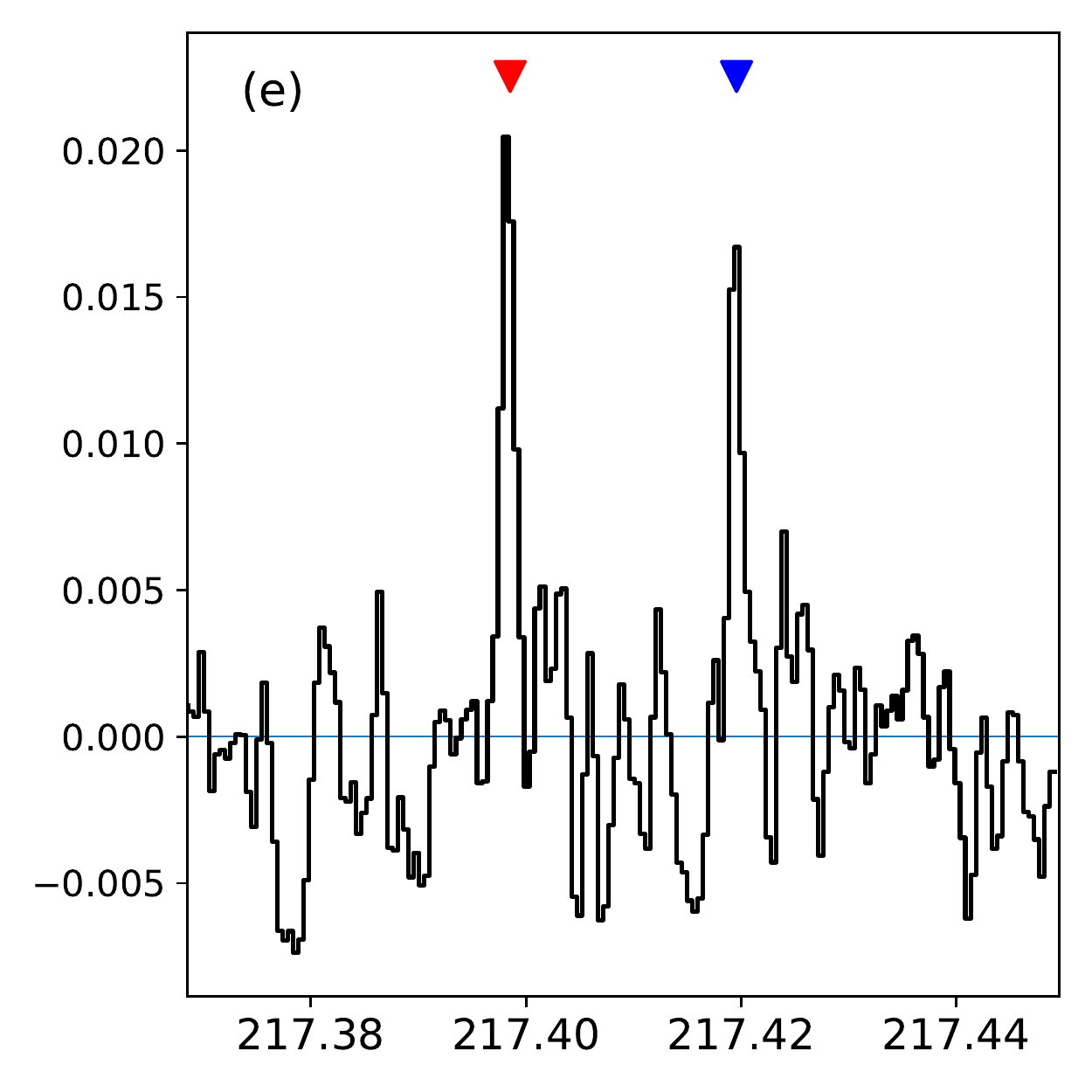}
\includegraphics[scale=0.3]{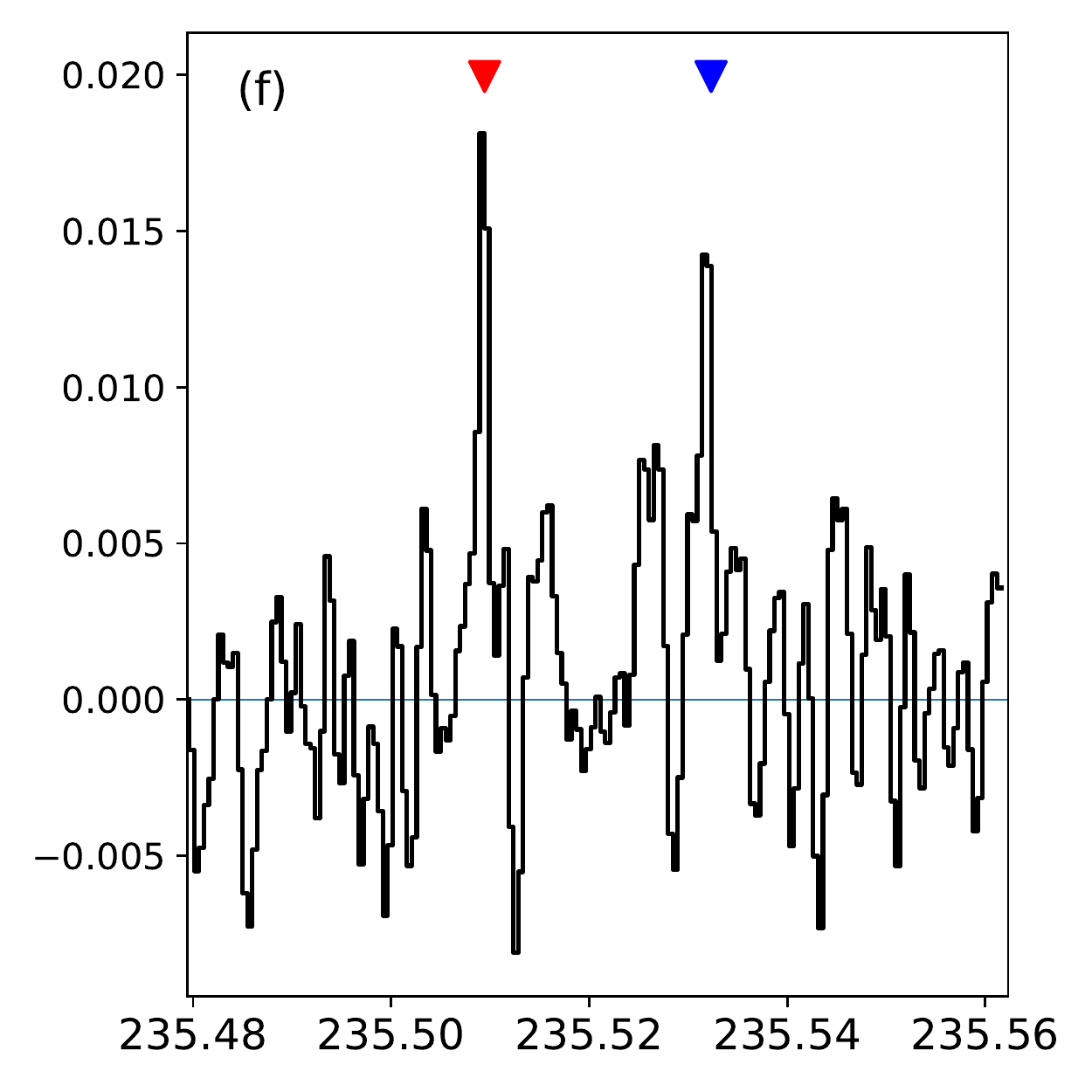}
\includegraphics[scale=0.3]{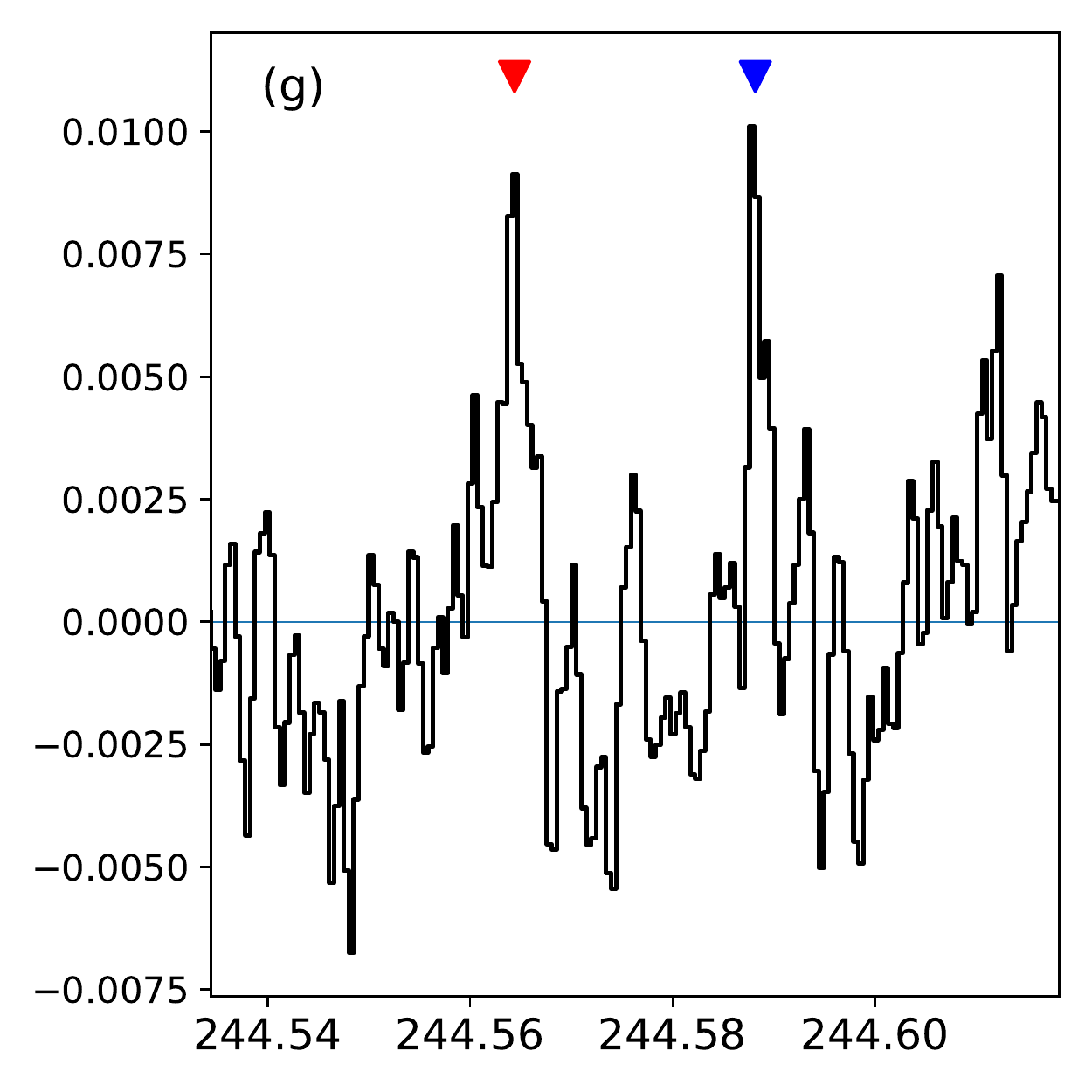}
\includegraphics[scale=0.3]{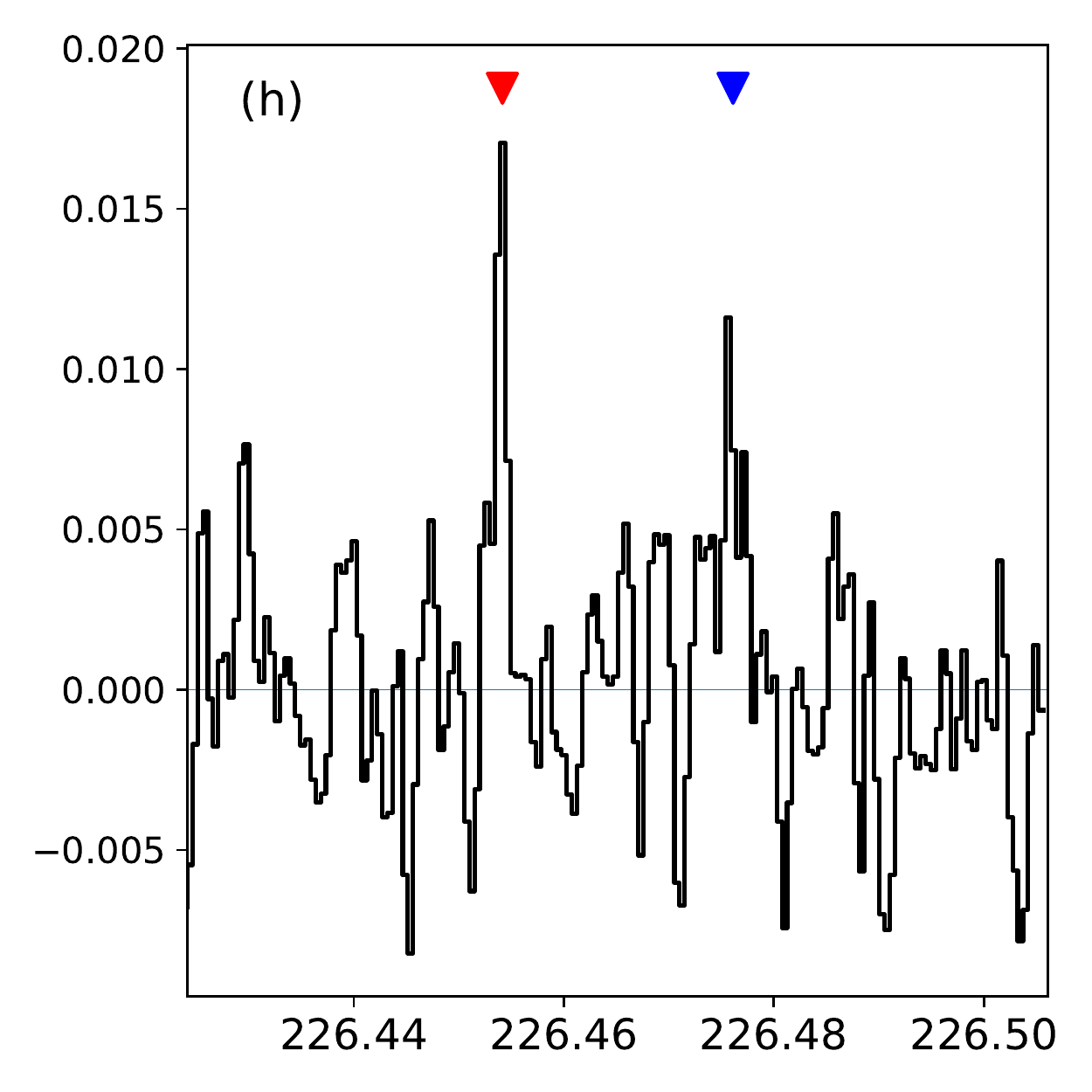}
\includegraphics[scale=0.3]{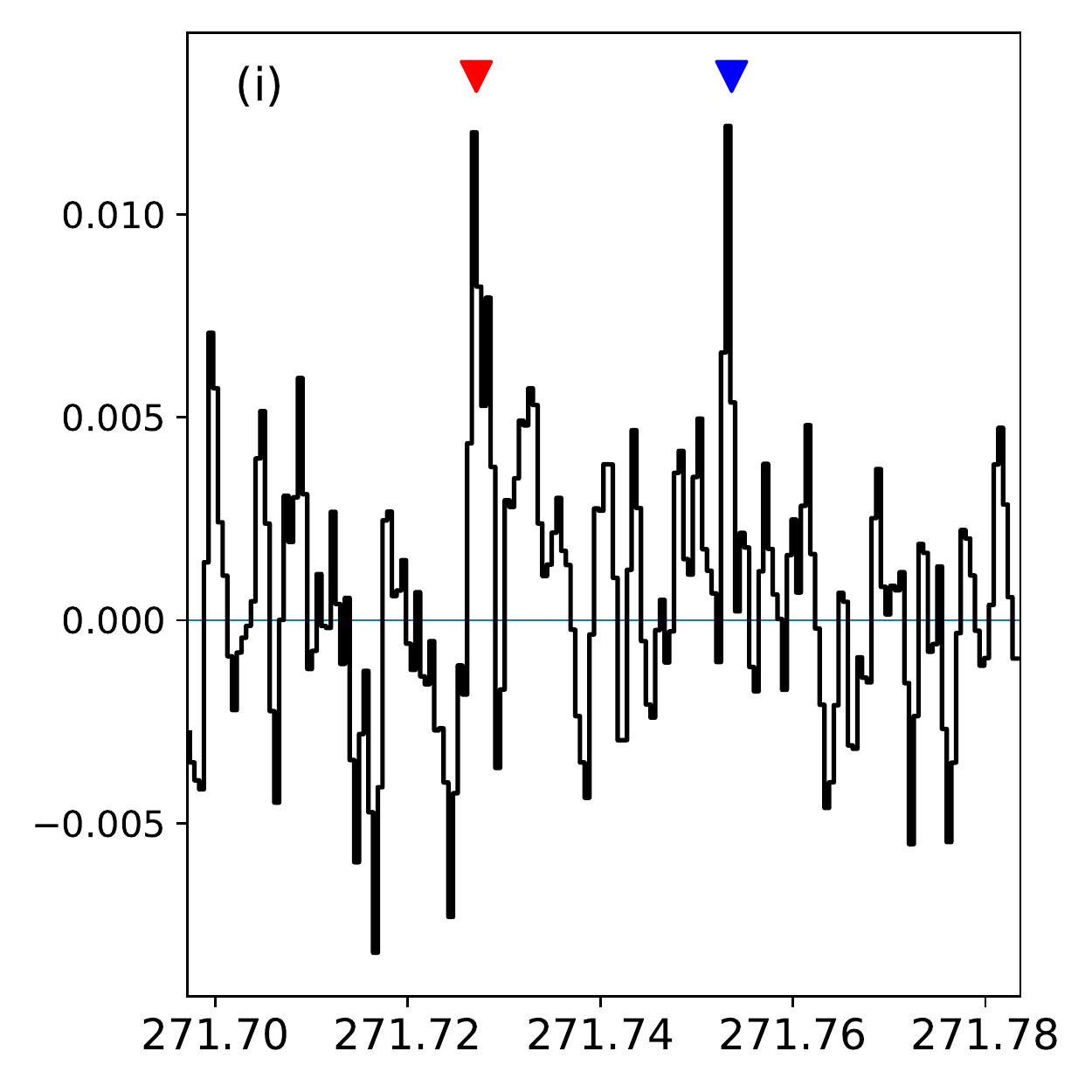}
\includegraphics[scale=0.3]{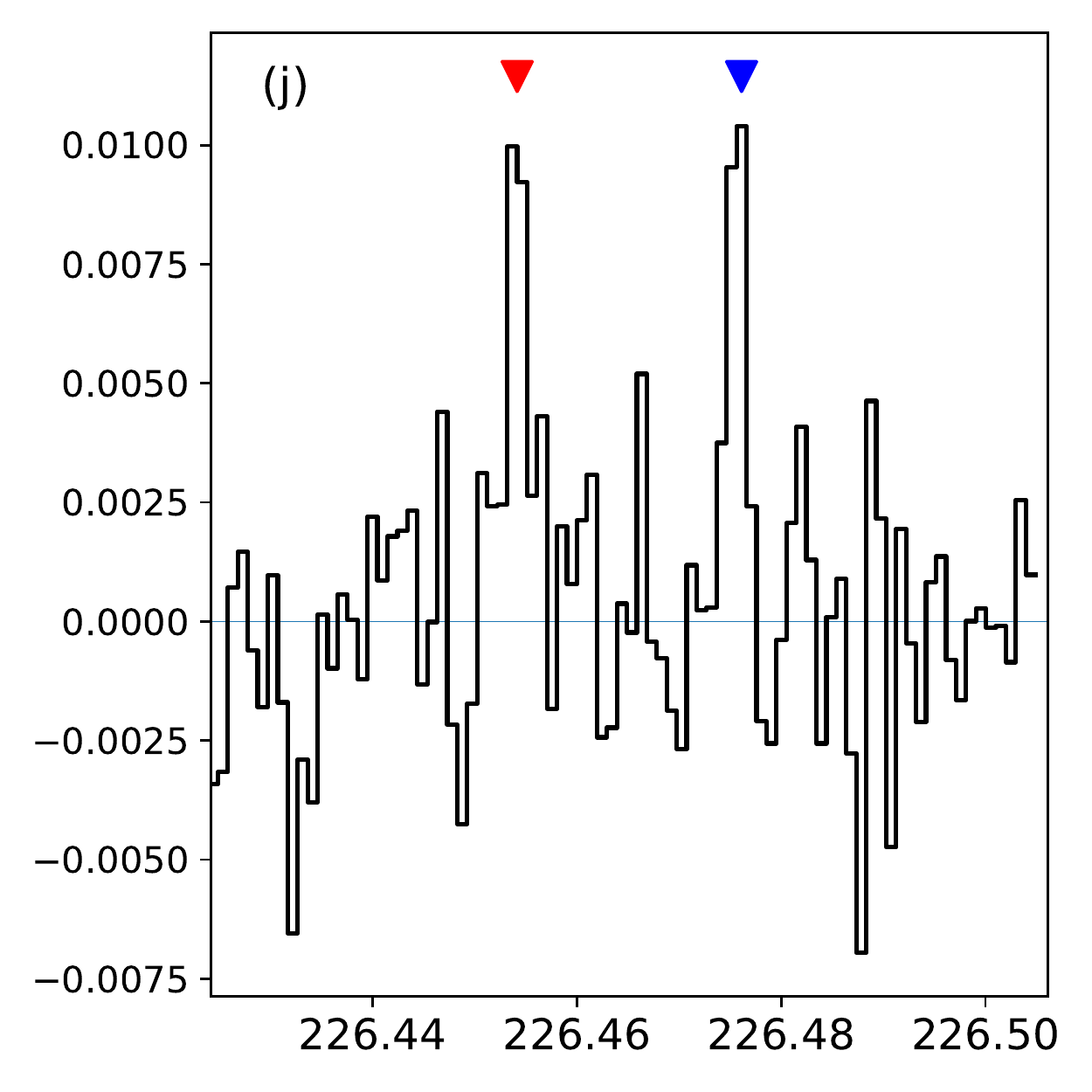}
\includegraphics[scale=0.3]{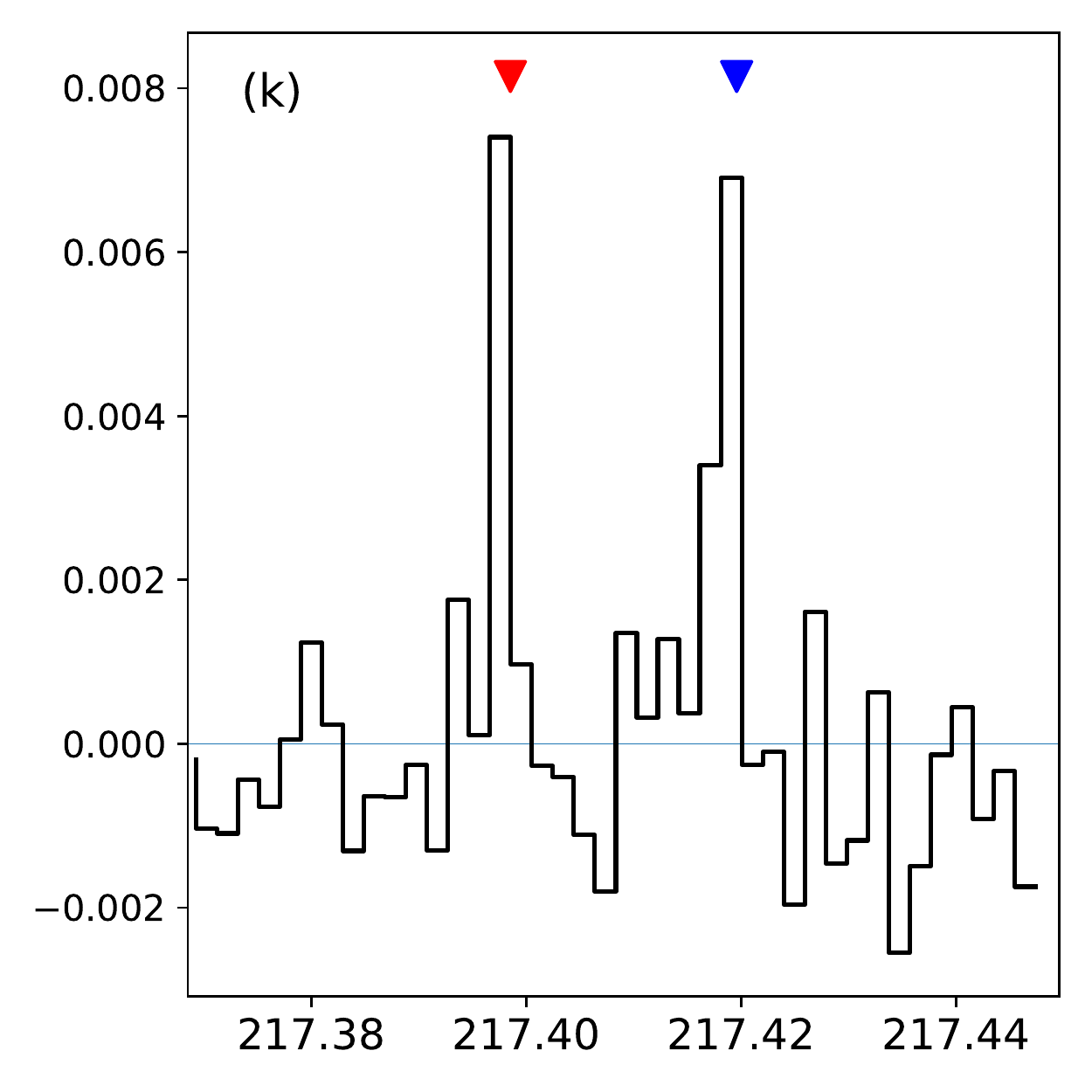}
\includegraphics[scale=0.3]{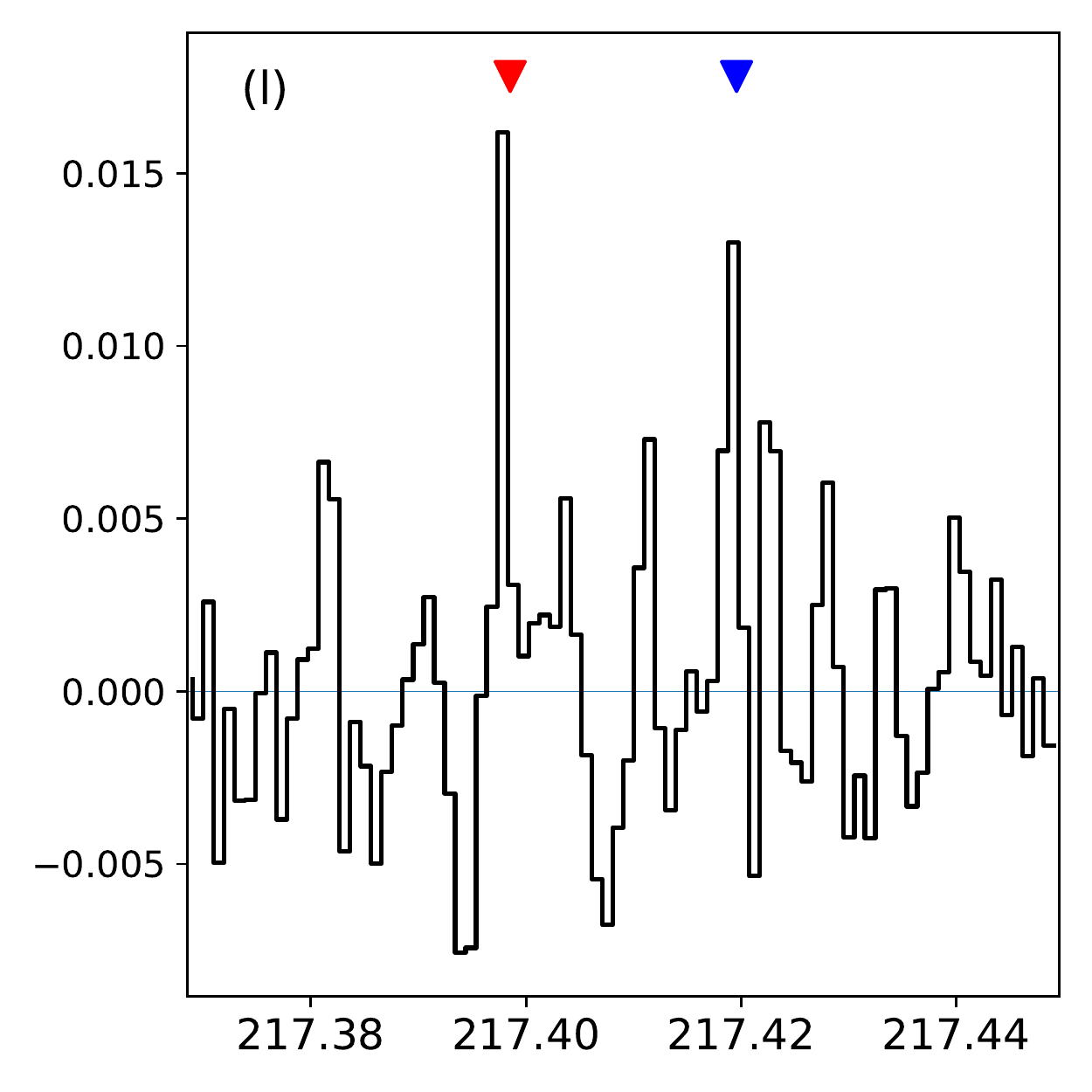}
\caption{Spectra of \ce{HC^{13}CCN}(red marker) and \ce{HCC^{13}CN}(blue marker). X and Y axes are for rest frequency (GHz) and intensity (Jy/beam), respectively. Order of panel is corresponding to Table \ref{table:observation} from upper left to bottom right. \label{fig:HC13CCN-HCC13CN-spectra}}
\end{center}
\end{figure*}

\begin{figure*}
\begin{center}
\includegraphics[scale=0.5]{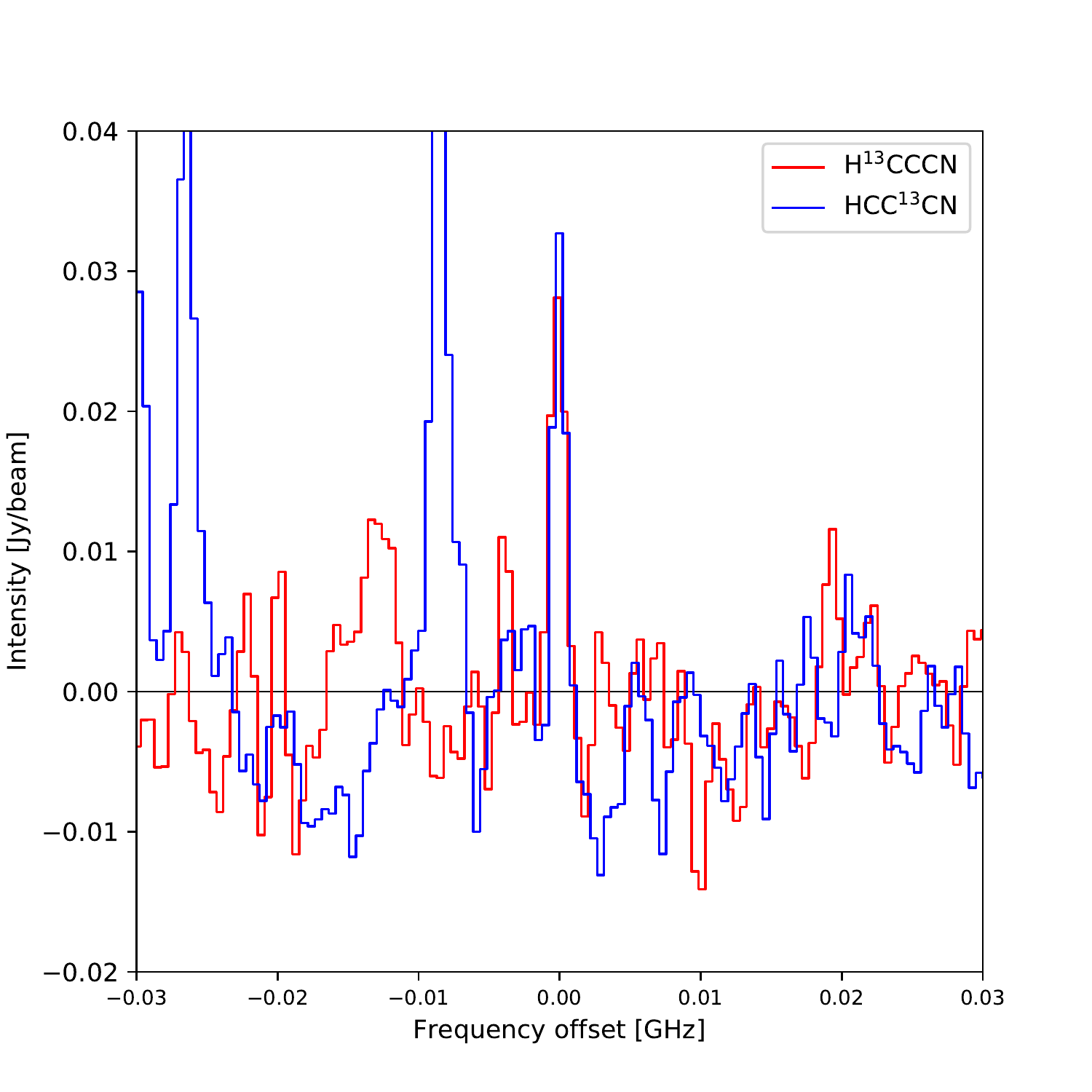}
\caption{Over-plotted spectra of \ce{H^{13}CCCN}($J$=35--34, red) and \ce{HCC^{13}CN}($J$=34--33, blue). Two strong \edit1{spectral lines} shown in the left of \ce{HCC^{13}CN} are \edit1{ $J$=36--35 transition group of \ce{C2H5CN}. At the left edge, a \ce{HC^{13}CCN }($J$ = 34--33) line which is blended with one of \ce{C2H5CN} line is present.} \label{fig:H13CCCN-HCC13CN-spectra}}
\end{center}
\end{figure*}

Figures \ref{fig:HC13CCN-HCC13CN-spectra} and \ref{fig:H13CCCN-HCC13CN-spectra} show the obtained spectra for each pair. 
The \ce{HC^{13}CCN} and \ce{HCC^{13}CN} lines are plotted in the same panels in Figure \ref{fig:HC13CCN-HCC13CN-spectra} due to small differences in frequency, while the lines for \ce{H^{13}CCCN} and \ce{HCC^{13}CN} are over-plotted in Figure \ref{fig:H13CCCN-HCC13CN-spectra}.
For \ce{H^{13}CCCN} including data, as shown in Figure \ref{fig:H13CCCN-HCC13CN-spectra}, since the \ce{HC^{13}CCN} line was blended with a \edit1{Ethyl Cyanide (\ce{C2H5CN(36_{1,36} - 35_{1,35}})) line}, only the \ce{HCC^{13}CN} line was used for the intensity comparison. 
\edit1{Note that the detection of \ce{C2H5CN} with ALMA was reported previously \citep{Cordiner2015}.}

\edit1{
In Titan's atmosphere, the chemical processes associated with \ce{HC3N} vary with altitude, with ion chemistry being dominant at high altitudes and neutral chemistry at low altitudes.
Previous ALMA observation study \citep{Thelen2019} derived an altitude range where \ce{HC3N} ($J$=35--34) is sensitive by the radiative transfer analysis of the disk averaged spectra.
The optically thick line core region of $\pm$ 2 MHz from the line center probes at the $\sim$800 km high altitude region, whereas wings has sensitivity at 150 km, where \ce{HC3N} shows abundance peaks in the high latitude regions.
Because the obtained intensities of isotopomer lines analyzed in our study are quite weaker than that of \ce{HC3N} line, we infer that the isotopomer lines have sensitivity at the lower stratosphere.
}

\subsection{Abundance ratio calculation}
To calculate the abundance ratios for the two pairs of isotopologues, as described methods below, we compared the measured integrated line intensities instead of retrieving the vertical abundance using radiative transfer method.
Since they share similar vertical abundance, optical depth and the same temperature profile, their relative abundance can be derived with a proper consideration of the difference of the rotational transitions between the two isotopologues. 
In addition, optically thin molecular lines enable us to assume that the measured intensity is proportional to the optical depth. 
This method was also applied to the previous [\ce{H^{13}CCCN}]:[\ce{HCCC^{15}N}] measurement using ALMA observation result \citep{Cordiner2018}. 
Because we did not need to consider the effect of the difference of beam size, temperature profile and three-Dimensional distribution of \ce{HC3N}, the only effect on the line intensity that was estimated and applied was the difference in spectroscopic parameters with respect to the stratospheric temperature. 
For the calculation of the parameters relating to opacity such as the partition function and population, the equations used were those listed in the appendix of \cite{Turner1991} and \cite{Iino2014}.
Line parameters such as the Einstein coefficient $A_{ul}$, the lower state energy $E_l$ and the rotational constant $B$, were obtained from the NASA JPL catalogue \citep{Pickett1998}. 
The considered excitation temperatures were from 140 to 180 K under the assumption of the local temperature equilibrium condition.

\begin{figure}
\begin{center}
\includegraphics[scale=0.5]{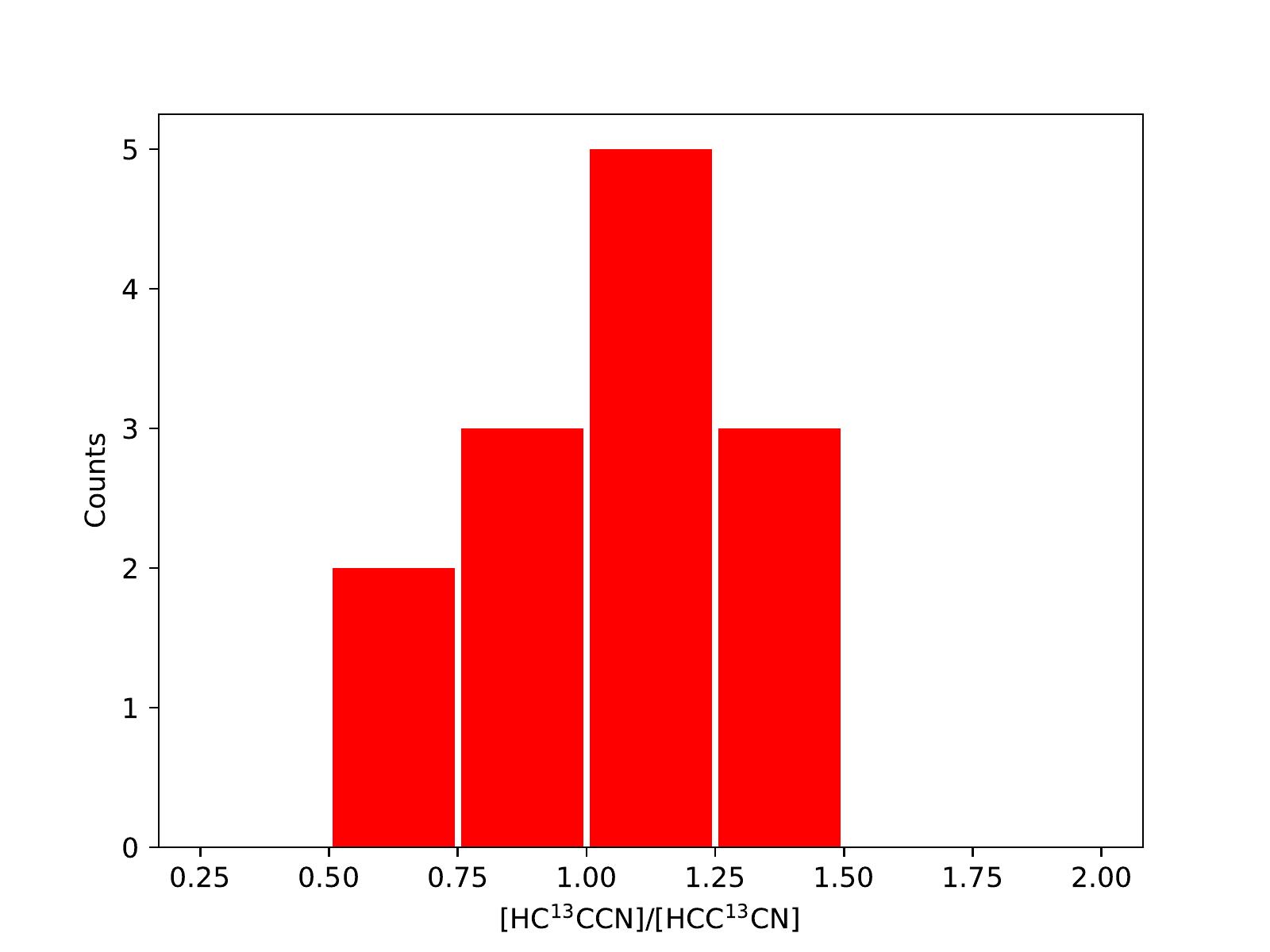}
\caption{A histogram of derived [\ce{HC^{13}CCN}]/[\ce{HCC^{13}CN}]. \label{fig:histogram} }
\end{center}
\end{figure}

For the \ce{HC^{13}CCN} -- \ce{HCC^{13}CN} pair, the evaluation was simple because they share the same rotational state $J$.
For the range of analyzed rotational transitions, the line intensity difference between \ce{HC^{13}CCN} and \ce{HCC^{13}CN} was estimated to be less than 0.02$\%$ for the modeled temperature range.
The only exception was the $J$=24--23 transition, where \ce{HCC^{13}CN} has three hyperfine splitting lines. 
In this case, the line intensities of three transitions were simply integrated.
As a result, the line intensity difference for the $J$=24--23 pair was determined to be $\sim$0.2$\%$.
Thus, considering the line intensity difference between two isotopologues and optically thin line intensities, for all the \ce{HC^{13}CCN} -- \ce{HCC^{13}CN} pair, we used the integrated line intensity ratios as the abundance ratio.
The derived [\ce{HC^{13}CCN}]/[\ce{HCC^{13}CN}] ratios are shown in the rightmost column of Table \ref{table:observation}. 
Figure \ref{fig:histogram} shows a histogram of the obtained [\ce{HC^{13}CCN}]/[\ce{HCC^{13}CN}] ratio.
Any fractionation relating to the rotational transition was found.
The averaged mean value is 1.09 with a standard deviation of 0.25.
\edit1{
Since the time variation of the isotopic ratios are beyond the scope of this paper, we simply average the data taken from 2012 to 2015 epoch. 
}

For the \ce{H^{13}CCCN} -- \ce{HCC^{13}CN} pair, taking into account the different rotational states, it was determined that \ce{HCC^{13}CN}($J$=34--33) has a 6.4$\%$ higher intensity than \ce{H^{13}CCCN}($J$=35--34) in average under the assumed excitation temperature range. 
With this correction, we found the abundance ratio of [\ce{H^{13}CCCN}]/[\ce{HCC^{13}CN}] to be 1.17$\pm$0.20. 
Note that this error was determined from the noise level of single spectrum.

\section{Discussion}
The measured [\ce{H^{13}CCCN}]:[\ce{HC^{13}CCN}]:[\ce{HCC^{13}CN}] ratios on Titan are nearly uniform within their errors.
The ratios do not show the large anomalies as reported by the ISM observations that \ce{HCC^{13}CN} and/or \ce{H^{13}CCCN} show 35--110 and 20$\%$ higher abundance than the others, respectively\citep{Takano1998, Araki2016,Taniguchi2016, Taniguchi2017}. 
Assuming that no time variation of $^{12}$C/$^{13}$C ratio is present since the previous observation, the result indicates that the $^{12}$C/$^{13}$C ratios on three isotopologues are same as \edit1{79$\pm$17} that measured on \ce{H^{13}CCCN} \citep{Jennings2008}. 
Below, the derived results are discussed from two viewpoints: $^{13}$C concentration on \ce{HCC^{13}CN} and \ce{H^{13}CCCN}. 
The obtained absence of carbon fractionation process on Titan is possibly explained by the environmental difference such as atmospheric temperature between Titan and the ISM, and may constrain chemical reactions present in Titan's middle and upper atmosphere.

\subsection{ \ce{HCC^{13}CN} concentration}
Assuming the main \ce{HC3N} production as Reaction \ref{eq:CN}, in case of \ce{HCC^{13}CN} concentration exists, $^{13}$C concentrations on CN and/or its precursor are greater than that in reaction counterparts, \ce{C2H2}, whose $^{12}$C/$^{13}$C ratio was determined to be 84.8$\pm$3.2 using the infrared spectra obtained by the Cassini spacecraft \citep{Nixon2008}. 
A main production pathway of CN is a photo-dissociation of HCN \citep{Loison2015}.
The other pathway, a photo-dissociation of \ce{C2N2}, is negligible because the abundance is below 1--0.1$\%$ of HCN.
The most recent ALMA observation reported that no significant $^{13}$C concentration in HCN (89.8$\pm$2.8: \cite{Molter2016}) in relative to \ce{C2H2}.
Thus, if exists, $^{13}$C concentration on CN occurs after the photolysis of HCN.

An ion-molecule isotope exchange process between \ce{^{13}C+} and CN has been proposed for exothermic $^{13}$CN concentration process for interstellar clouds as follows \citep{Colzi2020}:
\begin{equation}
    \ce{^{13}C+ + CN <-> C+ + ^{13}CN + \Delta E(31.1K)} \label{eq:13CN} 
\end{equation}
Reaction \ref{eq:13CN} likely causes \ce{HCC^{13}CN} enrichment in some ISM, especially in low-temperature conditions ($\sim10$ K) \citep{Takano1998}.
However, Reaction \ref{eq:13CN} does not seem to work effective on Titan's relatively higher atmospheric temperature (140 -- 180 K in the stratosphere) than interstellar, because the backward reaction of Reaction \ref{eq:13CN} can proceed and suppress the isotopic fractionation of CN in such high temperature environment.

The other scenario is relating to abundance of \ce{^{13}C+} ion. 
\cite{Vuitton2007} calculated \ce{C^{+}} number density as 1.4$\times$10$^{-2}$ cm$^{-3}$ considering mas spectral measurement result.
The calculated density is smaller than other major ions such as \ce{CH2+}, \ce{CH3+}, \ce{CH5+}, \ce{N+} and so on, thus the Reaction \ref{eq:13CN} may not work effective to concentrate $^{13}$C on CN and subsequently \ce{HC3N}.

\subsection{ \ce{H^{13}CCCN} concentration}
Reaction \ref{eq:HNC} is the only pathway to produce \ce{HC3N} from \ce{C2H} radical.
Because \ce{C2H2}, a precursor of \ce{C2H}, is a symmetric carbon molecule, an anomaly between \ce{H^{13}CCCN} and \ce{HC^{13}CCN} is caused by the abundance difference between \ce{C^{13}CH} and \ce{^{13}CCH}. 
In the ISM, the anomaly is possibly due to the isotope exchange reaction to achieve $^{13}$C concentration on \ce{C^{13}CH} as follows  \citep{Furuya2011}: 
\begin{equation}
    \ce{^{13}CCH + H <-> C^{13}CH + H +$\Delta$E( 8.1K)} \label{eq:CCH}
\end{equation}

Similar to the case of Reaction \ref{eq:13CN}, forward reaction of Reaction \ref{eq:CCH} is considered to be active only in low temperature environment such as the starless cores. 
In case of Titan, Reaction \ref{eq:CCH} is not expected to work for concentration of C$^{13}$CH due to the high temperature condition.
In addition, because of low HNC abundance \citep{Moreno2011, Cordiner2014}, contribution of the Reaction \ref{eq:HNC} for \ce{HC3N} production may be negligible. 

\section{Summary and future prospect}
We have detected the presence of all three $^{13}$C substituted species of \ce{HC3N}, namely, \ce{H^{13}CCCN}, \ce{HC^{13}CCN} and \ce{HCC^{13}CN}, in Titan's atmosphere using observational data from ALMA archive. 
The statistically derived [\ce{HC^{13}CCN}]/[\ce{HCC^{13}CN}] value was determined to be 1.09$\pm$0.25, whereas \edit1{that of measured in } starless dark clouds and low-mass star forming regions were previously reported to excess the present error. 
Additionally, [\ce{H^{13}CCCN}]/[\ce{HCC^{13}CN}] was found to be 1.17$\pm$0.20, although this result is less reliable than that for [\ce{HC^{13}CCN}]/[\ce{HCC^{13}CN}] because of a single pair detection. 
For both cases, no significant $^{13}$C concentration in \ce{HC3N} was detected, which differs from most of the ISM cases. 

\edit1{The large environmental difference with ISM is the high atmospheric temperature environment of Titan. 
A recent ALMA temperature measurement revealed that the measured stratospheric temperature above 100 km is at least 130 K, and reaches 180 K at 300 km \citep{Thelen2018}. 
These temperatures are quite higher than that expected in the ISM as 10 K \citep{Taniguchi2019}.
In such a cold region with temperatures around 10 K, the barrierless and exothermic isotopic exchange reactions, Reactions \ref{eq:13CN} and \ref{eq:CCH}, which have been considered to cause fractionation on \ce{HC3N}, are driven by the differences in the zero point energy.
On the other hand, in a high temperature environment, such as Titan's stratosphere and mesosphere, the backward reactions of Reactions \ref{eq:13CN} and \ref{eq:CCH} can proceed to suppress the isotopic fractionation in the precursors of \ce{HC3N} and \ce{HC3N} itself. 
}


Recently, similar to the case of Titan, \cite{Taniguchi2021} reported uniform carbon isotopic ratios of \ce{HC3N} around a massive young star objects.
They proposed that \ce{HC3N} is mainly produced via \ce{HC3NH+} ion, which has complicated formation pathway, which introduces more complicated pathways of \ce{HC3N} formation pathways and thus the entire \ce{HC3N} isotopic ratios would be affected by other reactions than Reactions \ref{eq:13CN} and \ref{eq:CCH}.
\edit1{
In addition to Reactions \ref{eq:CN} and \ref{eq:HNC}, a recent study suggested that the photo-dissociation of \ce{C2H3CN} and H-atom addition to \ce{HC4N2} may produce \ce{HC3N} \citep{Vuitton2019}. 
}
\edit1{
For the total understanding of Titan's isotopic fractionation processes, such ion-relating reactions and newly proposed neutral reactions should be investigated.
}

\edit1{
Similar to Reaction \ref{eq:13CN}, new exchange reactions of \ce{^{13}C} and \ce{^{13}C+} with C-bearing species are proposed by recent publications \citep{Colzi2020, Loison2020}. 
These newly proposed reactions may induce \ce{^{13}C} concentrations in C-bearing species, in particular if \ce{^{13}C} and \ce{^{13}C+} are abundant which is unlikely in Titan atmosphere. 
Our result, together with the previously reported non-concentration of \ce{^{13}C} in  C-bearing species on Titan such as \ce{CH4}, \ce{C2H2}, HCN, \ce{HC3N} and CO, may be interpreted as the consequence of \ce{^{13}C} and \ce{^{13}C+} depletion.
}

\acknowledgments
This study makes use of the ALMA data listed in the table \ref{table:observation}.  
ALMA is a partnership of ESO (representing its member states), NSF (USA) and NINS (Japan), together with NRC (Canada), MOST and ASIAA (Taiwan), and KASI (Republic of Korea), in cooperation with the Republic of Chile. The Joint ALMA Observatory is operated by ESO, AUI/NRAO and NAOJ. 
This work was supported by grants from the Telecommunications Advancement Foundation (TI), the Japan Society for the Promotion of Science (JSPS) Kakenhi (17K14420, 19K14782, 20K14523, 20K04046 and 20K04017) and the Astrobiology Center Program of National Institutes of Natural Sciences (NINS).

\newpage

\newpage
\begin{rotatetable}
\begin{deluxetable*}{ccccc|cc|cc|c}
\tabletypesize{\scriptsize}
\tablecaption{Summary of observation parameters and result with error  \label{table:observation}}
\tablehead{
& &  & &  & HC$^{13}$CCN &  & HCC$^{13}$CN & & \\
Project code & Obs. & Obs. & Beam & $\Delta$f & Freq. & Integrated  & Freq. & Integrated  & Abundance \\
& date & time & shape & &  &  intensity &  & intensity & Ratio \\
& y-m-d& (s) & ($\arcsec$)& (kHz) & (GHz) & (Jy/beam km/s)  &  (GHz) & (Jy/beam km/s)  &
}
\startdata
(a) 2011.0.00735&2012-01-13&217.68&2.00$\times$1.30&488&226.4541&0.0614(7.1)&226.4761&0.0712(8.2)&0.86(0.16)\\
(b) 2011.0.00735&2012-01-13&217.68&1.87$\times$1.21&488&244.5644&0.0451(4.2)&244.5882&0.0726(6.8)&0.62(0.17)\\
(c) 2012.1.00932&2014-03-11&157.344&1.15$\times$0.75&244&217.3985&0.0284(7.8)&217.4196&0.0196(5.4)&1.44(0.32)\\
(d) 2012.1.00248&2014-04-29&158.304&0.81$\times$0.59&488&235.5094&0.0188(5.9)&235.5323&0.0199(6.2)&0.95(0.22)\\
(e) 2012.1.00248&2014-04-29&158.304&0.88$\times$0.65&488&217.3985&0.0345(12.1)&217.4196&0.0251(8.8)&1.37(0.19)\\
(f) 2012.1.00248&2014-04-29&158.304&0.95$\times$0.58&488&235.5094&0.0255(6.8)&235.5323&0.0216(5.8)&1.18(0.27)\\
(g) 2012.1.00453&2014-07-07&157.344&0.48$\times$0.44&488&244.5644&0.0150(5.3)&244.5882&0.0125(4.1)&1.19(0.35)\\
(h) 2013.1.00271&2015-04-05&157.344&1.45$\times$1.01&488&226.4541&0.0214(6.3)&226.4761&0.0161(4.7)&1.34(0.35)\\
(i) 2012.1.00453&2015-05-13&157.344&0.66$\times$0.47&488&271.7271&0.0124(5.1)&271.7536&0.0104(4.3)&1.20(0.37)\\
(j) 2015.1.01312&2016-03-21&151.2&0.90$\times$0.72&976&226.4541&0.0191(5.1)&226.4761&0.0172(5.3)&1.11(0.32)\\
(k) 2015.1.00512&2016-03-31& \edit1{604.8} &1.07$\times$0.84&1953&217.3985&0.0105(4.6)&217.4196&0.0143(6.3)&0.73(0.20)\\
(l) 2015.1.00315&2016-04-02&755.0&1.09$\times$0.84&976&217.3985&0.0199(4.3)&217.4196&0.0201(4.4)&1.00(0.32)\\ \hline
& &  & &  & H$^{13}$CCCN &  & HCC$^{13}$CN & & \\
2012.1.00178&2015-04-22&157.344&1.19$\times$0.74&488&308.5041&0.0293(8.4)&307.9689&0.0264(6.5)&1.17(0.20)\\
\enddata
\tablecomments{Integrated intensities and abundance ratios are followed by 1-$\sigma$ S/N ratio in parenthesis. Abundance ratios are expressed as [\ce{HC^{13}CCN}]/[\ce{HCC^{13}CN}] or [\ce{H^{13}CCCN}]/[\ce{HCC^{13}CN}].}
\end{deluxetable*}
\end{rotatetable}

\newpage

\bibliography{library}
\end{document}